\documentclass[twocolumn]{aastex63}

\usepackage{tikz}
\usepackage{savesym}
\savesymbol{tablenum}
\usepackage{siunitx}
\restoresymbol{SIX}{tablenum}
\usepackage{graphicx}
\usepackage{float}
\hypersetup{%
  colorlinks=true,
  linkbordercolor=red,
  pdfborderstyle={/S/U/W 1}
}

\newcommand\hst{\emph{HST}}

\graphicspath{{./figures/}}

\received{XXXX}
\revised{YYYY}
\accepted{ZZZZ}

\submitjournal{ApJ}

\shorttitle{SED Modeling of NGC~4395}
\shortauthors{Cruz et al.}

\begin{document}

\title{Modeling the SED of the AGN inside NGC~4395}

\author[0000-0002-1775-3602]{Hector Afonso G. Cruz}
\affiliation{William H. Miller III Department of Physics and Astronomy, The Johns Hopkins University, Baltimore, MD 21218}
\author[0000-0003-4700-663X]{Andy D. Goulding}
\affiliation{Department of Astrophysical Sciences, Princeton University, Princeton, NJ 08540, USA}
\author[0000-0002-5612-3427]{Jenny E. Greene}
\affiliation{Department of Astrophysical Sciences, Princeton University, Princeton, NJ 08540, USA}

\begin{abstract}

We study the broad-band spectral energy distribution (SED) of the prototypical low-mass active galactic nucleus (AGN) in NGC~4395. We jointly model the optical through mid-infrared SED with a combination of galaxy and AGN light, and find that on arcsecond scales, the AGN dominates at most wavelengths. However, there is still some ambiguity about emission from the galaxy, owing partially to the strong short-term variability of the black hole. We investigate the use of smooth and clumpy-torus models in order to disentangle the nuclear infrared emission, as well as exploring the use of poloidal wind emission to account for the blue spectral slope observed in the near-IR. Even when simultaneously fitting the full optical--IR spectral range, we find that degeneracies still remain in the best-fit models. We conclude that high spatial resolution and wider wavelength coverage with the \emph{James Webb Space Telescope} is needed to understand the mid-infrared emission in this complex highly-variable object, which is the best nearby example to provide a blueprint to finding other low-mass AGN via their mid-infrared emission in the future.

\end{abstract}

\keywords{Active galactic nuclei (16), Intermediate-mass black holes (816), Spectral energy distribution (2129)}

\section{Introduction} \label{sec:intro}

Super-massive black holes are ubiquitous in the centers of massive galaxies, with mass ranges of $10^{6} - 10^{9} \, M_{\odot}$  \citep[e.g.,][]{kormendyho2013}. On the opposite extreme are stellar mass black holes exhibiting masses on the order of $10^{1} - 10^{2} M_{\odot}$, formed from the deaths of massive stars \citep{remillardmcclintock2006}. It is therefore reasonable to suspect that a further population of black holes should exist in the mass regime between both extremes, the ``intermediate-mass'' black holes (IMBHs). 

The number and mass distribution of IMBHs will depend on the physical mechanism that makes the seeds \citep[see][ and references therein]{volonteri2010,greeneetal2020}. Seed mechanisms include the collapse of Population III stars with 100~$M_{\odot}$ seeds \citep[e.g.,][]{fryeretal2001}, direct collapse to $10^4-10^5$~$M_{\odot}$ seeds \citep{brommloeb2003,loebrasio1994,lodatonatarajan2006,begelmanetal2006}, or gravitational runaway in dense star clusters to make $10^3$~$M_{\odot}$ seeds \citep[e.g.,][]{ebisuzakietal2001,millerhamilton2002}.

Those seeds that do not continue to grow beyond this point should leave behind relic IMBHs with $M_{\rm BH} \approx 10^{2} - 10^{5} \, M_{\odot}$, which should provide clues to their formation. To date, direct evidence for only one such IMBH has been found with $M_{\rm BH} \approx 150$~$M_{\odot}$ \citep{gwIMBH2020}. These are therefore prime science objectives of the next generation gravitational wave observatories such as \emph{LISA} \citep{amaro-seoaneetal2015}, sensitive to detecting the first black hole seeds out to redshifts $z \sim 20$ at masses $10^{4} - 10^{7} M_{\odot}$ to investigate SMBH formation at cosmic dawn \citep[e.g.,][]{bellovaryetal2019}.

In the interim, the challenge is to identify IMBH candidates using electromagnetic signatures. While possible dynamical detections of IMBHs in globular clusters have been reported \citep[e.g.,][]{gebhardtetal2005,lutzgendorfetal2013}, in every case there are contradictory masses in the literature \citep[e.g.,][]{tremouetal2018}, highlighting how challenging it will be to detect such objects should they exist. Beyond the Local Group, there are a handful of dynamical detections of $M_{\rm BH} \approx 10^5 M_{\odot}$ BHs \citep{nguyenetal2018,nguyenetal2019}. To reach statistical samples still requires looking for signatures of accretion, as has been attempted with optical spectroscopy \citep[e.g.,][]{reinesetal2013,moranetal2014}, X-ray \citep{milleretal2015,sheetal2017,pardoetal2016}, and optical variability \citep{baldassareetal2018}. 

\begin{deluxetable*}{lccDCDCCC}
\tablenum{1}
\tablecaption{A list of observations used in this study. \label{tab:observations}}
\tablewidth{0pt}
\tablehead{
\colhead{Instrument} &  \colhead{Channel} &  \colhead{Filter} &  \multicolumn2c{Pivot $\lambda$ } &  \colhead{Band $\Delta\lambda$} &  \multicolumn2c{Plate Scale} &  \colhead{R} &  \colhead{Exp. Time} &  \colhead{FOV} \\
\colhead{} &  \colhead{} &  \colhead{} &  \multicolumn2c{$[\si{\micro\meter}]$} &  \colhead{$[\si{\micro\meter}]$} &  \multicolumn2c{[arcsec/pix]} &  \colhead{$\left[\frac{\lambda}{\Delta\lambda}\right]$} &  \colhead{[sec]} &  \colhead{$[\mathrm{arcsec\times arcsec}]$}
}
\decimals
\startdata
ZTF & - & $g$ & 0.480 & 0.409 - 0.552  & 1.01  & - & 30 & 47 \; \si{deg^{2}} \\
ZTF & - & $r$ & 0.624 & 0.560 - 0.732  & 1.01  & - & 30 & 47 \; \si{deg^{2}}\\
ZTF & - & $i$ & 0.766 & 0.703 - 0.888  & 1.01  & - & 30 &  47 \; \si{deg^{2}}\\
\hline 
SDSS & SDSS-II &  & 0.681 & 0.3812 - 0.9815    & 0.396  & 4337 \lesssim R \lesssim 4348 & 53.9 & 3'' \si{diameter} \\
\hline
\hst/WFC3 & UVIS & F275W & 0.271 & 0.0165 & 0.04 & - & 4 \times 380 & 162 \times 162 \\
\hst/WFC3 & UVIS & F336W & 0.336 & 0.0158 & 0.04 & - & 4 \times 290 & 162 \times 162 \\
\hst/WFC3 & UVIS & F438W & 0.433 & 0.0197 & 0.04 & - & 4 \times 115 & 162 \times 162 \\
\hst/WFC3 & UVIS & F547M & 0.545 & 0.0296 & 0.04 & - & 4 \times 95  & 162 \times 162 \\
\hst/WFC3 & UVIS & F814W & 0.803 & 0.0663 & 0.04 & - & 4 \times 75  & 162 \times 162 \\
\hst/WFC3 & IR & F127M & 1.274 & 0.0249 & 0.13 & - & 4 \times 60  & 123 \times 136 \\
\hst/WFC3 & IR & F153M & 1.532 & 0.0379 & 0.13 & - & 4 \times 60  & 123 \times 136 \\
\hline
MAGNUM   & - & $K$ & 2.205 & 0.26   & 0.28 & - & 1080 & 2.4 \\
\hline
Spitzer & IRAC & 3.6 \si{\micro\meter} & 3.550 & 0.750  & 0.6  & - & 26.8 & 1711.8 \times 1821.0 \\
Spitzer & IRAC & 4.5 \si{\micro\meter} & 4.493 & 1.015  & 0.6  & - & 26.8 & 1711.8 \times 1821.0 \\
Spitzer & IRS  & SL2 & 6.36 & 5.12 - 7.60  & 1.8  & 60 \lesssim R \lesssim 127 & 121.9 & 3.6 \times 57.0 \\
Spitzer & IRS  & SL1 & 10.875 & 7.46 - 14.29  & 1.8  & 61 \lesssim R \lesssim 120 & 121.9 & 3.7 \times 57.0 \\
Spitzer & IRS  & LL2 & 17.585 & 13.90 - 21.27  & 5.1  & 57 \lesssim R \lesssim 126 & 243.8 & 10.5 \times 168.0 \\
Spitzer & IRS  & LL1 & 29.905 & 19.91 - 39.90  & 5.1  & 58 \lesssim R \lesssim 112 & 243.8 & 10.7 \times 168.0 \\
\enddata
\end{deluxetable*}

Additional information may come from focusing on the rest-frame infrared (IR) emission from putative low-mass BHs. High-ionization mid-IR emission lines that are relatively insensitive to dust obscuration and host-galaxy dilution effects are very effective at identifying active galactic nuclei (AGN) arising from low mass BHs  \citep{satyapaletal2007,gouldingalexander2009}. The mid-IR continuum from AGN is dominated by emission from a dusty ``torus'' of gas and dust that absorbs UV light from the accretion disk and re-emits in the infrared. 

Over the past decade, adaptive optics and interferometry have provided a new level of understanding of the torus region. We now appreciate that smooth torus models cannot simultaneously fit the spectral shape and Si absorption of AGN \citep[e.g.,][]{netzer2015}; we will further confirm this finding here even for a low-mass LLAGN such as NGC~4395. Furthermore, high resolution imaging shows that the torus comprises at least two components, a disk-like and a poloidal component. A compelling possibility is that the poloidal component arises from a wide-angle outflow or wind component \citep[e.g.,][]{honigetal2017}. However, disk-wind models of the torus have not been extended to low black hole mass before. Both for the purposes of identifying new low-mass black holes, and for understanding the geometry and dependence on physical parameters, it is thus crucial to model the tori of lower-mass systems. We make a start on this goal here with the AGN in NGC~4395.

The central IMBH powering the AGN at the heart of NGC 4395, a Type I Seyfert galaxy $\sim 4 \,\si{Mpc}$ \citep{thimetal2004}, is one of the nearest and best-studied IMBH candidates in a galaxy nucleus. NGC 4395 houses a relatively low luminosity AGN \citep{filippenkosargent1989,filippenkoho2003} with $L_{bol} \sim 10^{40}$~erg/s \citep{petersonetal2005}, and a current BH mass estimate of $3.5 \times 10^{5} M_{\odot}$ \citep{denbroketal2015}, although \citet{wooetal2019} posit the mass to be much lower at $\sim 10^{4} M_{\odot}$. Perhaps due to its low mass and/or low luminosity \citep{elitzurho2009}, this AGN is one of the most variable known \citep{moranetal2005}, varying at X-ray energies by a factor $\sim$3 on 2--3 hour timescales \citep{kammounetal2019}.

The goal of this paper is to compile and investigate the near-UV to MIR spectral energy distribution (SED) of the central region of NGC 4395, and model the photometry with a combination of templates representing the host galaxy, the accretion disk, and the dusty torus, to yield insight into low-luminosity AGN architecture. In \S \ref{sec:data} we introduce all the data sets and apertures that we use, in \S \ref{sec:sedfit} we present the broad-band fits with clumpy torus models, and in \S \ref{sec:discussion} we put NGC 4395 in the context of other samples of AGN with fitted torus parameters, and summarize our conclusions.

\begin{figure*}[t!]
    \centering
    \includegraphics[width=0.48\textwidth]{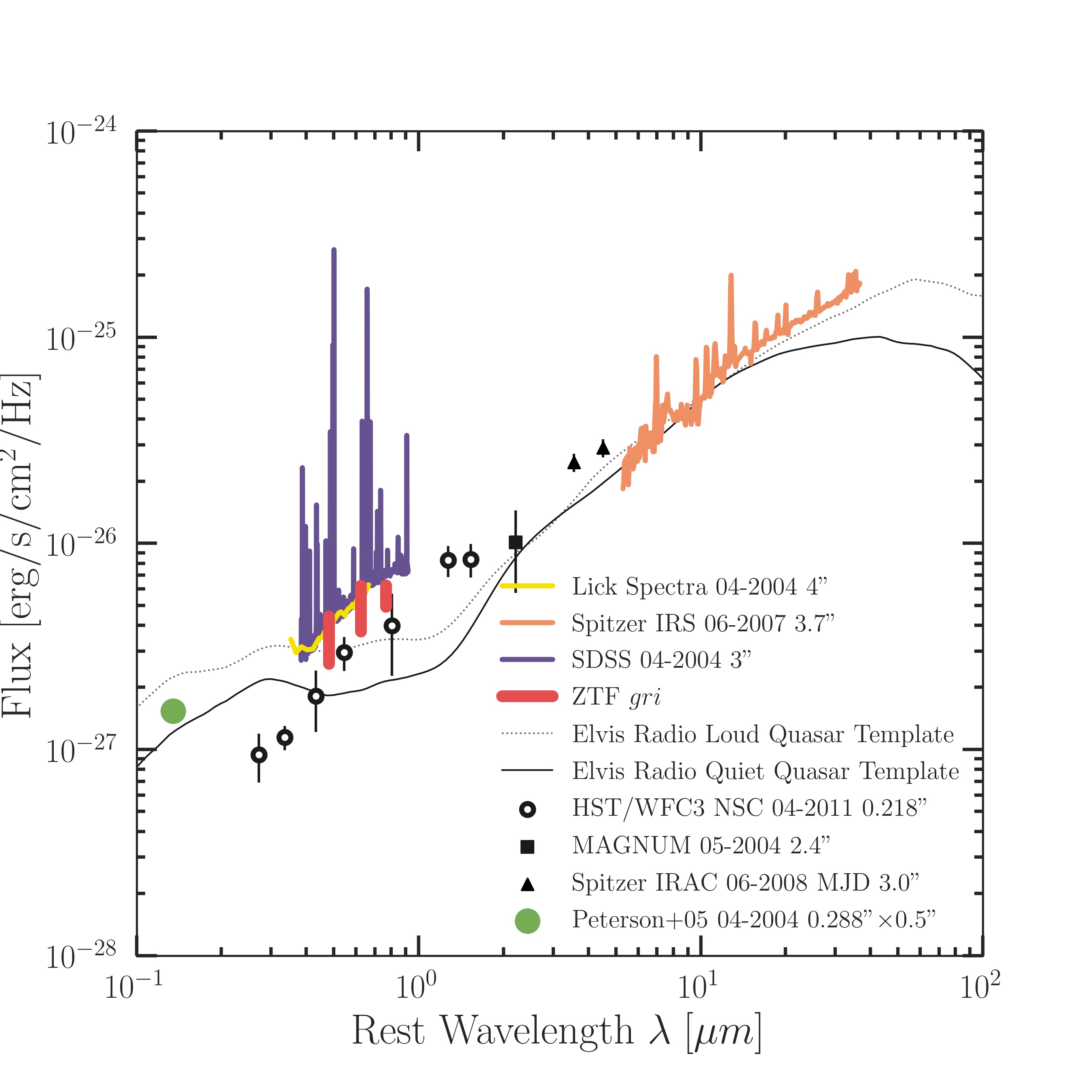}
    \includegraphics[width=0.48\textwidth]{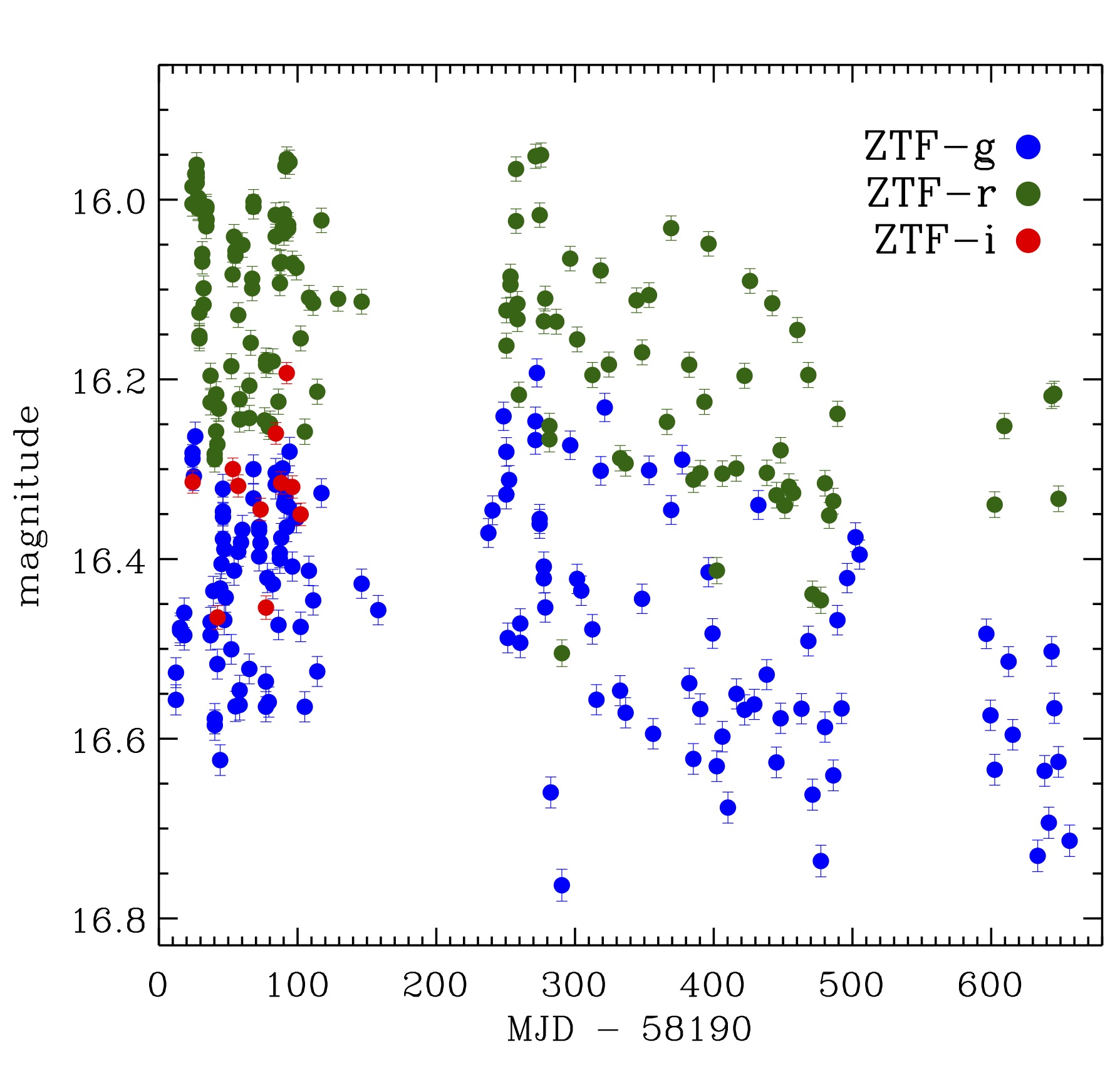}
    \caption{{\bf Left:} A compilation of various observations of NGC 4395, listed with corresponding apertures and dates of observation. We include published data from the \textit{Lick Observatory} \citep{desrochesetal2006}, the \textit{HST/WFC3} NSC photometry from \citet{carsonetal2015}, and ultraviolet data from \citet{petersonetal2005},  in addition to the archival SDSS, Spitzer IRS, and MAGNUM K-band data used in this study. To yield insight into the intrinsic optical variability of the AGN, the minimum and maximum $gri$-band photometry from the \textit{Zwicky Transient Facility} (ZTF) have been overplotted as red bars. For reference, \citet{elvisetal1994} quasar templates normalized to the Spitzer IRS spectrum at 12 \si{\micro\meter} have been overplotted. {\bf Right: } ZTF point source photometry for NGC~4395 in the $g$-, $r$- and $i$-bands (blue, green, red points, respectively). The day-long UV/optical variability is similar in both the $g$ and $r$ bands with $\sim$0.48~mags (AB).}
  \label{fig:timeSED}
\end{figure*}

\section{The Broad-band SED of NGC~4395} \label{sec:data}

To model the combined galaxy and AGN SED of NGC 4395, we make use of photometric and spectroscopic datasets covering the near-UV to the mid-IR (0.2--38.0$\mu m$; see Table~\ref{tab:observations}). We revisit the previously published SED from \citet{moranetal1999} with a set of new observations. There are two significant challenges in modeling the full SED of such a low-luminosity AGN. The first is that there is an unknown contribution from galaxy light that is likely both wavelength and aperture dependent. The second is that the AGN varies on significantly different timescales across the UV--IR wavelength range, making non-simultaneous SEDs challenging to interpret \citep[e.g.,][]{vaughanetal2005,burkeetal2020}. We confront these challenges directly in \S \ref{sec:variabilityandapertures}, but describe here various choices that we made to mitigate aperture mismatch.

At the highest spatial resolution, we harness broadband imaging from the UVIS and IR channels of the Wide Field Camera 3 on the \textit{Hubble Space Telescope (HST)} along with near-infrared imaging from the ground to pin down the likely contribution from galaxy light within our modeled aperture (\S \ref{sec:galaxysed}). Specifically, NGC 4395 harbors a nuclear star cluster (NSC) at its center \citep[e.g.,][]{carsonetal2015} that contributes some fraction of the light at all wavelengths. Using the \emph{HST} image, \citet{carsonetal2015} have spatially decomposed the contributions from the NSC and non-thermal AGN light in the optical/NIR wavelengths, giving us our best handle on the relative contributions of each component.

Over the full wavelength range, we also consider spectroscopy from the Infrared Spectrograph (IRS) on the \textit{Spitzer} Space Telescope (previously published in \citealt{hoodetal2017}) and the Sloan Digital Sky Survey (SDSS), as well as photometric measurements from the IRAC camera on Spitzer. 

In this section we present salient details about each data set that we include in the model.

\begin{table}[ht!]
\tablenum{2}
\caption{A list of photometry used for model SED fitting. \label{tab:photometry}}
\begin{center}
\begin{tabular}{lcccc}
\hline \hline
\multicolumn{1}{c}{Filter Name} &
\multicolumn{1}{c}{Pivot $\lambda$} &
\multicolumn{1}{c}{Band $\Delta\lambda$} &
\multicolumn{1}{c}{Flux Density} &
\multicolumn{1}{c}{Error} \\
\multicolumn{1}{c}{} &
\multicolumn{1}{c}{($\mu m$)} &
\multicolumn{1}{c}{($\mu m$)} &
\multicolumn{1}{c}{(mJy)} &
\multicolumn{1}{c}{(mJy)} \\
\hline
F275W & 0.271 & 0.1063 & 0.094 & 0.0252 \\ 
F336W & 0.336 & 0.0865 & 0.115 & 0.0154 \\ 
F438W & 0.433 & 0.1040 & 0.181 & 0.0598 \\ 
F547M & 0.545 & 0.1101 & 0.296 & 0.0556 \\ 
F814W & 0.810 & 0.3048 & 0.398 & 0.1702 \\ 
F127M & 1.274 & 0.8741 & 0.828 & 0.1431 \\ 
F153M & 1.532 & 0.9003 & 0.837 & 0.1556 \\
K-band & 2.205 & 0.48 & 1.659 & 0.4339 \\ 
\hline
SDSS 01 & 0.384 & 0.006 & 0.287 & 0.023 \\ 
SDSS 02 & 0.410 & 0.0469 & 0.311 & 0.022 \\ 
SDSS 03 & 0.460 & 0.0516 & 0.379 & 0.03 \\ 
SDSS 04 & 0.564 & 0.1283 & 0.491 & 0.057 \\ 
SDSS 05 & 0.643 & 0.0235 & 0.592 & 0.024 \\ 
SDSS 06 & 0.665 & 0.0132 & 0.662 & 0.029 \\ 
SDSS 07 & 0.769 & 0.0399 & 0.707 & 0.024 \\ 
SDSS 08 & 0.809 & 0.0399 & 0.731 & 0.021 \\ 
SDSS 09 & 0.849 & 0.0399 & 0.765 & 0.057 \\ 
SDSS 10 & 0.884 & 0.0299 & 0.724 & 0.041 \\ 
IRAC 1 & 3.556 & 1.175 & 2.468 & 0.247 \\ 
IRAC 2 & 4.501 & 1.59 & 2.905 & 0.291 \\ 
IRS 01 & 5.354 & 0.7 & 2.408 & 0.252 \\ 
IRS 02 & 6.311 & 1.2 & 3.017 & 0.348 \\ 
IRS 03 & 7.509 & 1.2 & 3.919 & 0.389 \\ 
IRS 04 & 8.708 & 1.2 & 4.134 & 0.255 \\ 
IRS 05 & 9.907 & 1.2 & 4.632 & 0.35 \\ 
IRS 06 & 11.106 & 1.2 & 5.85 & 0.176 \\ 
IRS 07 & 12.305 & 1.2 & 7.07 & 0.493 \\ 
IRS 08 & 13.505 & 1.2 & 7.929 & 0.537 \\ 
IRS 09 & 14.704 & 1.2 & 8.365 & 0.276 \\ 
IRS 10 & 15.853 & 1.1 & 9.048 & 0.066 \\ 
IRS 11 & 17.053 & 1.1 & 9.66 & 0.111 \\ 
IRS 12 & 18.253 & 1.1 & 10.119 & 0.119 \\ 
IRS 13 & 20.061 & 2.3 & 10.848 & 0.27 \\ 
IRS 14 & 22.511 & 2.4 & 12.093 & 0.148 \\ 
IRS 15 & 24.859 & 2.3 & 12.897 & 0.201 \\ 
IRS 16 & 27.258 & 2.3 & 13.928 & 0.066 \\ 
IRS 17 & 29.658 & 2.3 & 14.764 & 0.188 \\ 
IRS 18 & 32.057 & 2.3 & 15.85 & 0.317 \\ 
IRS 19 & 34.507 & 2.4 & 16.963 & 0.095 \\ 
IRS 20 & 36.856 & 2.3 & 18.053 & 0.399 \\ 

\hline \hline

\end{tabular}
\end{center}
\end{table}

\subsection{Spitzer IRAC Data}\label{sec:spitzerIRAC}
    
We use data from the cryogenic mission of the Spitzer Infrared Array Camera (IRAC), observed through IRAC channels 1 and 2 (Ch1, Ch2 respectively) with central wavelengths of 3.6 and 4.5 \si{\micro\meter} respectively. These observations (Program ID: 40204) were performed in Cycle 7. Individual frames (exposure times $\sim 26.8$ \si{\sec}) were mosaicked using MOPEX and were processed and calibrated using the IRAC Pipeline; cryogenic data were calibrated with IRAC Pipeline S18.18.0. Mosaicking resulted in a new pixel scale of $\sim 0.6'' \times 0.6''$, for a combined field of view of approximately $30.4' \times 28.5'$.
    
Aperture photometry was performed on the 3.6 and 4.5 \si{\micro\meter} IRAC data. Here we choose not to perform direct PSF fitting as point sources in the mosaicked IRAC images ($\sim 0.6''$/pixel) are undersampled. Photometry was conducted using the PhotUtils v0.6 package in Astropy v3.1.1. We extracted surface brightness profiles in Ch1 and Ch2 centered at the position of the AGN. To closely match other available multiwavelength data, we extract photometry from within a 3$''$ aperture. The observed surface brightness profile extends significantly beyond that expected from a simple PSF, which we model here as a simple Gaussian. Within  $r\sim1.5''$ (5 pixel diameter), the enclosed flux is higher than that expected from a point source. At the distance of NGC 4395, the AGN component is a point source, but the surrounding NSC is well-resolved \citep[e.g.,][]{carsonetal2015}. We find that $\sim$33\% of the light is therefore likely to come from stars in the NSC. The statistical uncertainties on these measurements were determined to be significantly smaller than the systematic uncertainty ($\sim$10\%) expected from performing photometry on mosaicked Spitzer images (see the IRAC Handbook), hence, we conservatively adopt $\sim$10\% uncertainties on each of these IRAC measurements. 

NGC~4395 has been observed on four separate occasions in Ch1 and 2 during the cold and warm phases of the Spitzer mission. Across the $\approx$6 year baseline, the typical variability was only $\sim$8\% from the mean in both Ch1 and Ch2, with the maximal amplitude of variability being observed in Ch2 of $\sim21\%$ across the whole ranges of observations. Hence, in the case of the IRAC observations, AGN variability is captured within our assumed photometric uncertainties.
    
\subsection{Spitzer IRS Spectra} \label{sec:spitzerIRS}

To extend the resolution and wavelength range of our analyses, we further include archival low-resolution (R $\sim$ 60 -- 127) mid-IR spectra (5.2--38.0$\mu$m) from Spitzer-IRS included in our fitting. The co-added and pre-processed, background subtracted, and calibrated data were retrieved from the NASA/IPAC Infrared Science Archive (IRSA) archive. The data had been processed with IRS Pipeline S18.18.0. To account for the different aperture sizes, the individual spectral orders from the Short-Low (SL) and Long-Low (LL) modules were matched using the overlapping wavelength coverage between the orders. The width of the SL slit (3.7$''$) is similar to the aperture size used in our IRAC photometry, and hence, we anchor the aperture corrections to the flux in the 2nd SL spectral order.
    
Directly mixing spectroscopy and photometry in SED fitting can produce statistically incoherent results due to the overweighting of the spectral elements in the fit. Hence, we artificially lower the resolution of the Spitzer-IRS data and produce synthetic photometry that captures the main features and spectral shape seen in the IRS data, such as Polycyclic Aromatic Hydrocarbons and silicates in absorption/emission. We separate the combined Spitzer-IRS spectrum into 20 synthetic top-hat filters with widths $\delta \lambda = 1.2 \mu$m in the range 5.7--18.9$\mu$m and 2.4$\mu$m in the range 18.9--38.1$\mu$m. Uncertainties are estimated from the RMS of the spectral values within the synthetic filter.

\subsection{Sloan Digital Sky Survey Spectrophotometry}\label{sec:sdss}
    
Our mid-IR spectrophotometric measurements produced from the Spitzer-IRS data may include contributions from both the central AGN and dust re-emission of starlight that resides within the 3$''$-width slit. To encompass similar AGN+stellar contributions at optical wavelengths, we include the well-matched 3$''$ fiber spectroscopy available in the 7th Data Release of the SDSS \citep{abazajianetal2009}. 
    
Following our methodology for the Spitzer-IRS spectroscopy, we construct synthetic spectrophotometry from the DR7 spectrum using constant transmission bandpasses in ten spectral regions. We avoid prominent emission lines that are not fitted by continuum models. Uncertainties were calculated from the standard deviation of the flux density of the individual spectral elements in the bandpass. See Table~\ref{tab:photometry} for the central wavelengths, bandpass widths and spectrophotometric measurements in each filter.
    
    \begin{figure*}[t!]
    \centering
    \includegraphics[width=\textwidth]{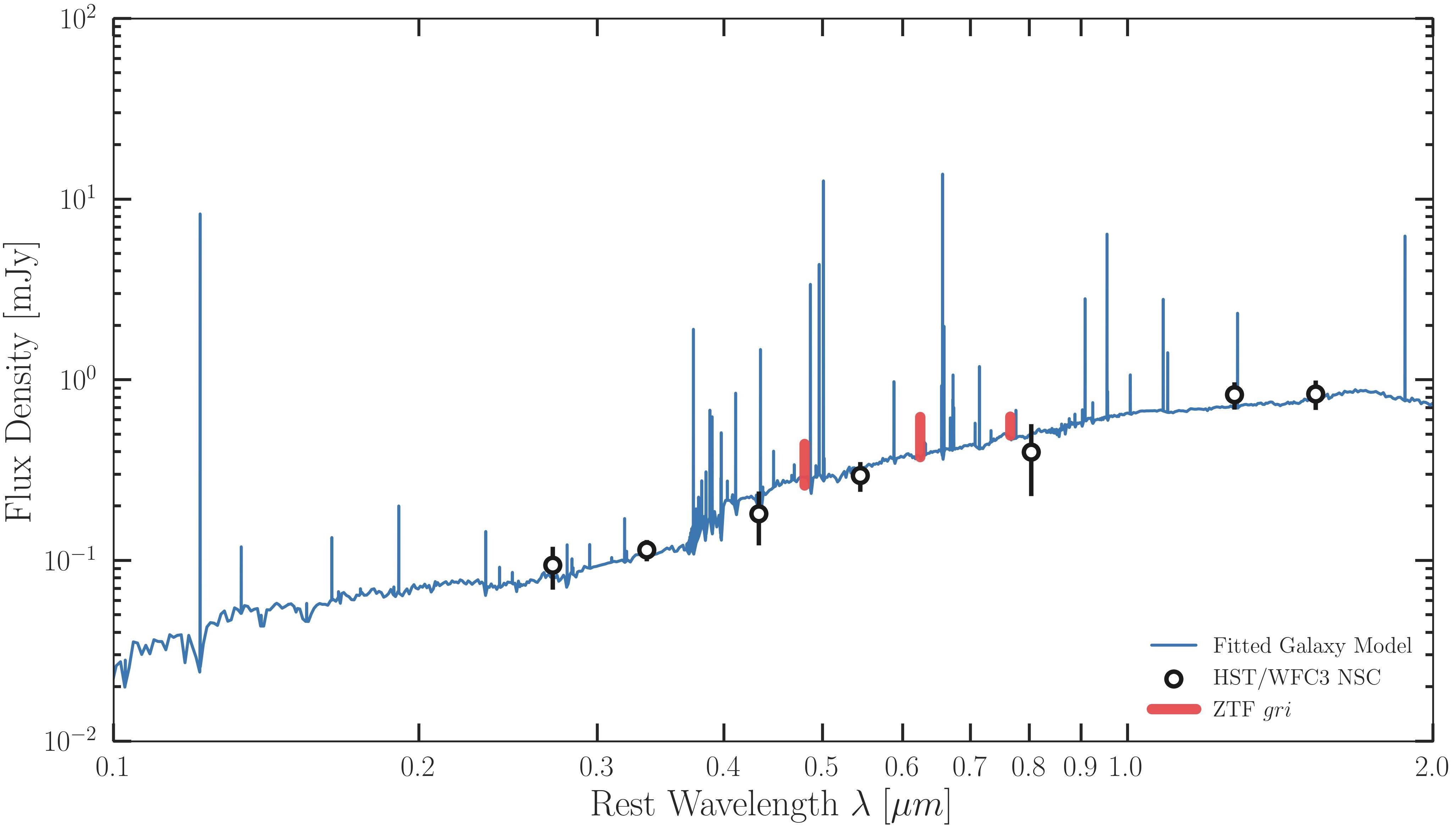}
    \caption{The best-fit stellar-only model SED, fitted using the \textit{HST/WFC3} NSC photometry from \citet{carsonetal2015}. We overplot the minimum and maximum archival \textit{ZTF} photometry, which we argue represents a minimum and maximum state of the AGN in NGC 4395. The \textit{HST} photometry for the NSC are consistent with the minimum \textit{ZTF} points, giving reassurance that the photometry representing the NSC have negligible AGN contamination.}
  \label{fig:fwdmodelSED}
\end{figure*}

\subsection{HST WFC3/UVIS Photometry}\label{sec:hst}

\hst\ provides the highest-resolution look at the AGN+NSC combination in the center of NGC 4395. We make use of 5-band \hst\ Wide Field Camera 3 (WFC3)-UVIS photometry from \citet{carsonetal2015} in the F275W, F336W, F438W, F547M and F814W filters. \citet{carsonetal2015} use the \hst\ imaging to construct surface brightness profiles of the galaxy center, and use {\tt\string GALFIT}  \citep{pengetal2002,pengetal2010} to fit a combined point source and Sersic profile \citep{sersic1963} to simultaneously model the emission from the AGN and the NSC. \citet{carsonetal2015} model the \emph{HST} data used in this paper to find an F814W effective radius of 4.56 \si{pc} ($\sim0.2$''). We anchor our model of the galaxy contribution to the SED using the photometry of the NSC, which dominates the galaxy light on the 3\arcsec\ scales used to construct the broad-band SED.

\subsection{Near-IR Photometry} \label{sec:magnum}

In the near-IR, we use \hst\ WFC3-IR (F127M; F153M; Carson et al. 2015) and ground-based K-band ($2.2 \, \si{\micro\meter}$) photometry (FWHM$\sim$1.1$''$) from the Multicolor Imaging Photometer (MIP) mounted on the 2m telescope of the Multicolor Active Galactic NUclei Monitoring (MAGNUM) project at the Haleakala Observatories in Hawaii \citep{minezakietal2006}.

We take the mean of these photometric data, and to encompass the variability seen in these measurements into our SED modeling, we take their standard deviation as the measurement ``uncertainty''.

\subsection{Zwicky Transient Facility Photometry}

Constraints on optical variability can be deduced using data from the \textit{Zwicky Transient Facility}, measured through \textit{gri} bands spanning a wavelength range of approximately 4000 - 9000 \AA ) and represent observations over a span of 694, 624, and 78 nights respectively, encompassing March 15 2018--January 15 2020. We note that \citet{burkeetal2020} also use the Transiting Exoplanet Survey Satellite (TESS) to measure optical variability in NGC 4395 on timescales $<$ one month.

\section{Apertures \& Variability} \label{sec:variabilityandapertures}

As mentioned above, the contribution from starlight is challenging to disentangle from AGN light, especially given the low-luminosity of the AGN in NGC 4395. Our overall approach is therefore to jointly model the AGN+galaxy light within a uniform aperture of $\sim 3\arcsec$. The width of the SL slit on the Spitzer IRS (3.7$''$) is similar to the fiber width used in recording our SDSS spectrophotometry of 3 $''$, and hence, we anchor the aperture corrections to the flux in the 2nd SL spectral order.

An even greater challenge, that cannot be fully addressed with the current data, is the variability of the AGN. NGC 4395 is one of the most variable AGN known in the X-ray \citep[e.g.,][]{vaughanetal2005}, and it has been observed to vary on timescales of minutes to years in the UV, optical, and NIR bands \citep[e.g.,][]{minezakietal2006,peterson2014,burkeetal2020}. Ideally, we would like a simultaneous measurement of the source across all wavelengths, to model the source in the same luminosity state. However, that is currently unavailable, and so in what follows we attempt to bracket the range of possible flux densities in bands with multi-epoch observations.

There have been studies of the optical through NIR variability of NGC 4395 that provide some insight into its variability amplitude. \citet{minezakietal2006} recorded intra-night J, K, and H band variations on the order of $\lesssim 10 \%$, with larger flux variations in optical to near-IR bands across days to months. Specifically, the K-band point used in our study, recorded 11 times over 345 days, was observed to fluctuate between $1.663 \pm 0.434 \; \si{mJy}$. Using \textit{ZTF}, Figure \ref{fig:timeSED} demonstrates that the minimum and maximum \textit{ZTF} points differ by a factor of about 1.6. More notably, the minimum \textit{ZTF} points on record are consistent with the measured \textit{HST} NSC photometry from \citet{carsonetal2015} on comparable spatial scales. Our interpretation is that the lowest flux points in the ZTF data represent a low state of the AGN when the optical light is actually galaxy-dominated, providing additional confirmation of our assumed galaxy fluxes as well as the range of AGN luminosities we might expect.

We will assume that the mid-IR varies on scales longer than a decade, supported by the lack of variability that we see in the few epochs of IRAC imaging (\S \ref{sec:data}). The torus surrounding the AGN is thought to extend to parsec scales, translating to a very long light-crossing time, and therefore long mid-IR variability time scale. However, given the low-mass nature of NGC 4395, we of course do not know whether the mid-IR variability timescale is also commensurately short, as suggested by the correlation between MIR to optical lag and luminosity \citep{yangetal2020}. We present the above observations in Figure \ref{fig:timeSED} along with their respective dates of observation.

\section{SED Fitting} \label{sec:sedfit}

We now turn to the main task of the paper, to jointly fit the AGN and galaxy light. To this end, we use the X-ray module adaptation to the Python-based Code for Investigating GALaxy Emission {\tt\string (X-CIGALE)} \citep{yangetal2020, nolletal2009, serraetal2011} to self-consistently model the contributions to the UV--IR SED of NGC 4395 from the galaxy star formation history, dust, AGN accretion disk, and torus. Specifically, we use the stellar population models from \citet{bruzualcharlot2003} combined with a \citet{chabrier2003} initial mass function and a delayed exponentially declining star-formation history. These optical/UV models are combined with a \citet{calzetti2000} attenuation law, which is balanced by dust re-emission in the IR following the empirical models of \citet{daleetal2014}. The free parameters associated with each model component are summarized in Table \ref{tab:fullCIG}.

Motivated by prior studies of AGN showing that the torus is inhomogenous \citep{mullaneyetal2011, netzer2015,mosheandisaac2006, nenkovaetal2008a, nenkovaetal2008, krolikandbegelman1988}, we parameterize the AGN component of the SED with {\tt skirtor} AGN models \citep{stalevskietal2012, stalevskietal2016, campsandbaes2015}. The {\tt skirtor} model assumes a clumpy geometry, parameterized by the average edge-on optical depth of the disk $t$, the radial power-law exponent that governs dust density $p_{l}$, the angular filling factor $q$, the half opening angle $\theta$, the ratio of the maximum to mininum radii $R$, and the inclination $i$. Through a radiative transfer model, the output SED consists of two components, primary emission from the disk and an anisotropic dust component (Table \ref{tab:fullCIG}). We will compare the relative successes of clumpy models over smooth ones in \S \ref{sec:smooth}. Furthermore, in \S~4.4, we will show that while the clumpy torus model shows dramatic improvements over the smooth torus models previously implemented by CIGALE, there are still aspects of the SED, in particular the 3-6 \micron slope, that are not well-modeled, leading us to also explore clumpy torus models with a disk and wind component \citep{honigetal2017}.

To mitigate degeneracies between the AGN and galaxy contributions to the SED (\S \ref{sec:variabilityandapertures}) we proceed in two steps. First, we place empirical limits on the galaxy SED using our highest spatial resolution (\emph{HST}+MAGNUM) data, for which the AGN and NSC components have been modeled separately (\S \ref{sec:variabilityandapertures}). Second, we take the allowed range of stellar continuum parameters from the initial galaxy-only fit, and refine the fit by simultaneously modeling the AGN and host galaxy components over our full spectral baseline, subject to the constraint that the galaxy component must provide a good fit to the \emph{HST} data.

\subsection{Nuclear Star Cluster SED Fitting} \label{sec:galaxysed}

The presence of an NSC at the center of NGC 4395 \citep{filippenkoho2003} translates into a non-negligible stellar contribution to the overall SED within 3\arcsec\ (\S \ref{sec:variabilityandapertures}). We first fit the \textit{HST} photometry with X-CIGALE, fitting only for the galaxy and dust parameters, and turning off the AGN component entirely. We emphasize that the NSC luminosity from \citet{carsonetal2015} is consistent with the low end of the \textit{ZTF} photometry, suggesting this is a reasonable assumed level for galaxy light on this spatial scale.

A plot of the best-fit SED model is shown in Figure \ref{fig:fwdmodelSED}. The model falls satisfactorily within the \textit{HST} errorbars, with a $\chi^{2}$ value of 2.2 with 7 degrees of freedom. This fit exhibits a bolometric dust luminosity of $\sim 4.9 \times 10^{39}$~erg/s and an unabsorbed stellar luminosity of $1.6 \times 10^{40}$~erg/s. The best-fit parameters are tabulated in Table \ref{tab:fullCIG} and suggest that the NSC emission is dominated by a weakly absorbed ($A_V \sim 0.1$) older ($\sim$9~Gyr old) stellar population with no additional evidence for a recent burst of star formation. These results are supported by the presence of a prominent 4000\AA\ break in the SED. Our approach in the rest of this section is to use this fit as a measure of true galaxy (dust + stellar emission) in NGC 4395. As there is degeneracy between AGN and galaxy light within the SDSS+\emph{Spitzer}/IRS, in our analyses we limit the fitting range for the stellar component such that this component continues to reproduce the \emph{HST} NSC data as described below. 

    \begin{figure}[t!]
    \centering
    \includegraphics[width=0.49\textwidth]{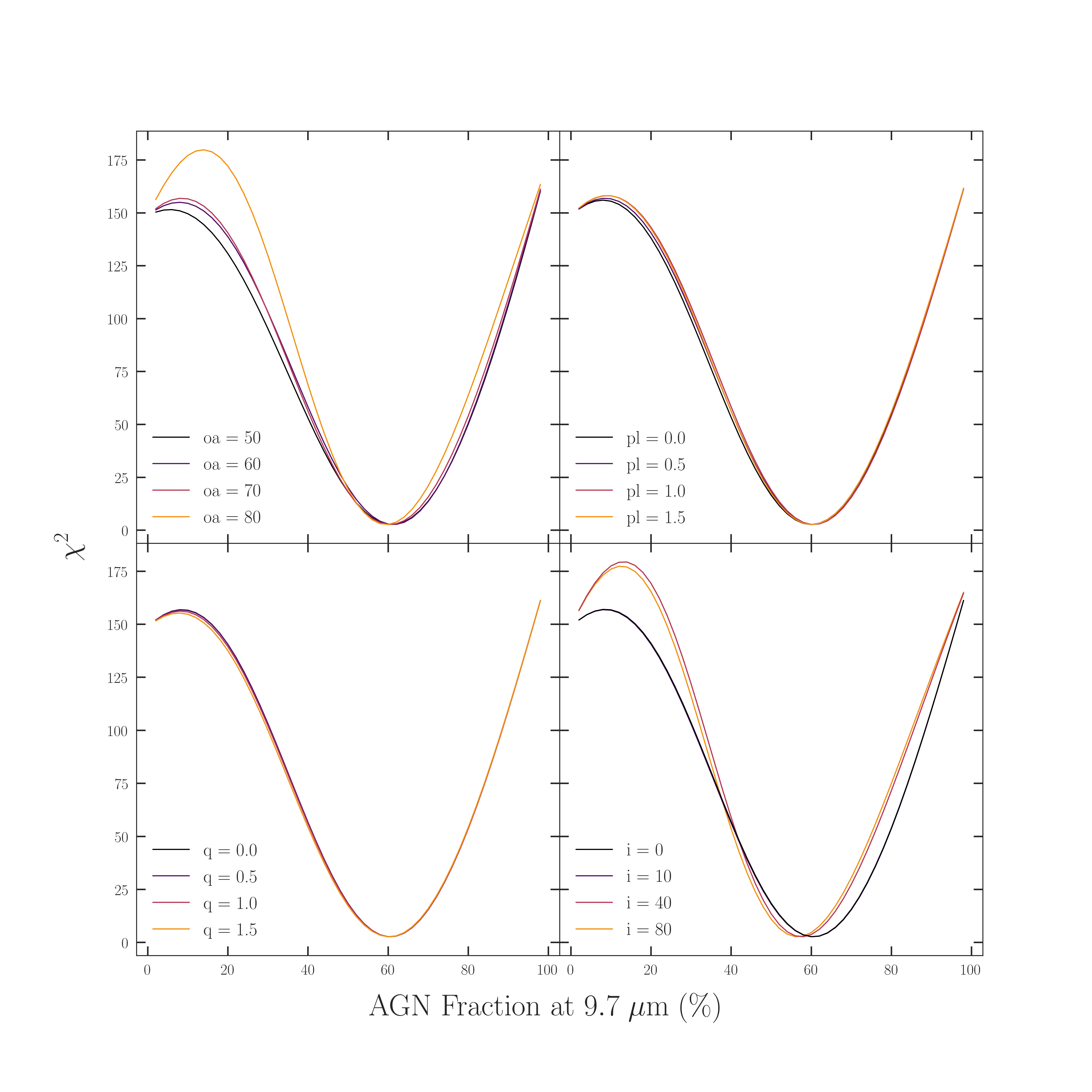}
    \vspace{-1.0cm}
    \caption{The $\chi^{2}$ of the \textit{HST/WFC3} photometry to the fitted galaxy model plotted against the fraction of the fitted AGN model at 9.7 \si{\micro\meter}, shown for different values for four of the most relevant {\tt skirtor} torus parameters. We seek the minimum $\chi^{2}$ value to ensure that the best-fit parameters that shape the global model SED in Figure \ref{fig:fullSED} encompass a model galaxy in satisfactory agreement with the NSC photometry. All models exhibit minimum $\chi^{2}$ values at an AGN fraction of approximately 60 \%, suggesting that the best-fit parameterization of the torus is in sufficient agreement with the global SDSS + IRS photometry as well as the galactic anchor specified in \S \ref{sec:galaxysed}.   }
  \label{fig:1dchi2}
\end{figure}

\begin{figure*}[t!]
    \centering
    \includegraphics[width=\textwidth]{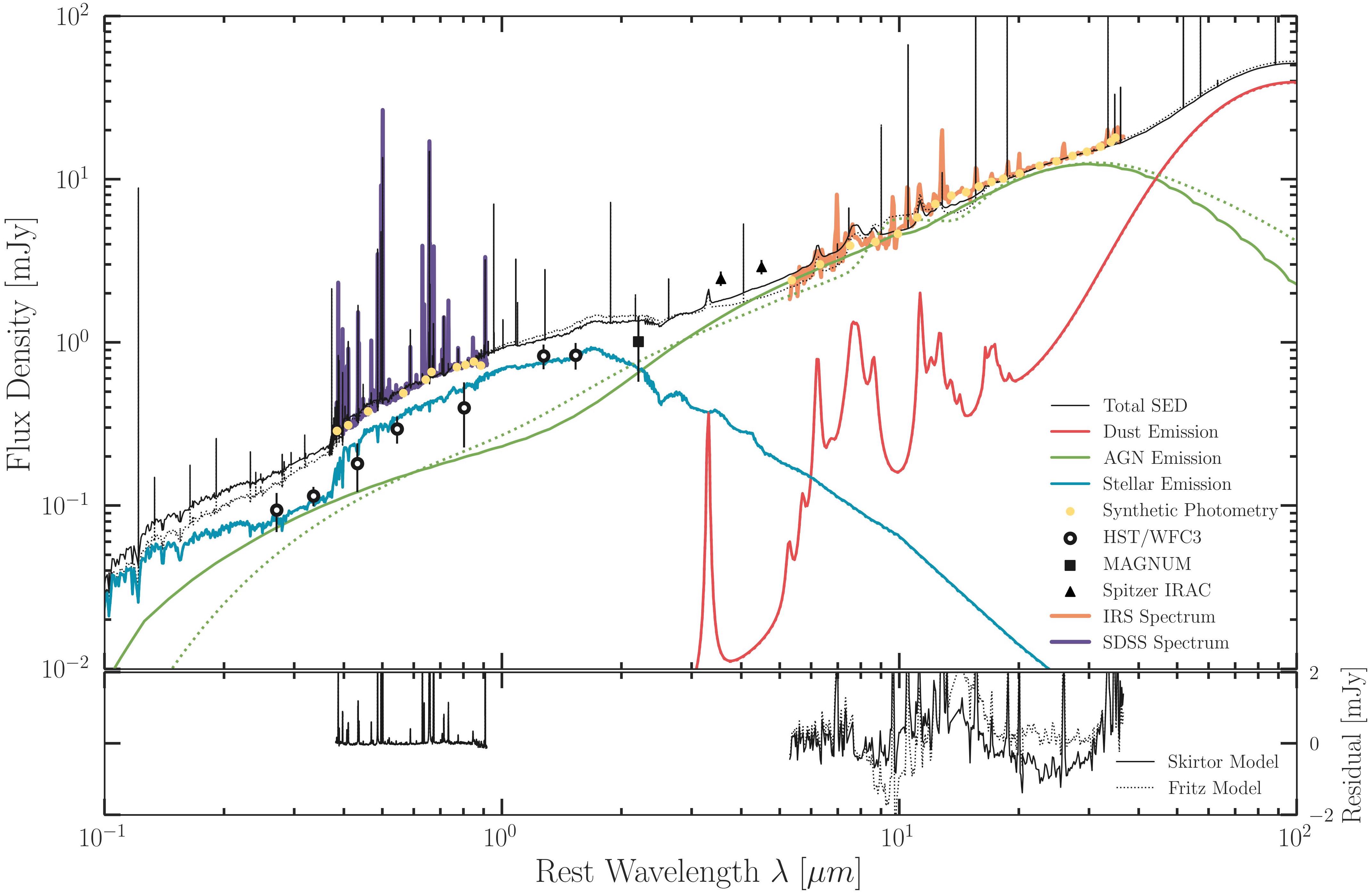}
    \caption{X-CIGALE SED modeling of the {\it Spitzer}-IRAC photometry (black triangles) and SDSS and {\it Spitzer}-IRS spectrophotometry described in \S\ref{sec:data}. Total best-fit SED, UV/optical stellar component, dust stellar component and smooth torus model are shown with black, blue, red and green lines, respectively. For illustration purposes only, overlaid are the $HST$ and MAGNUM high spatial resolution photometric data described in \S\ref{sec:hst} and \ref{sec:magnum}, as well the SDSS and {\it Spitzer}-IRS spectra. These points used in the forward-modeling process fit the data with a reduced $\chi^{2} = 2.71$.}
  \label{fig:fullSED}
\end{figure*}

\begin{deluxetable*}{rLCR}
\tablenum{3}
\tablecaption{A list of inputs and outputs from CIGALE for our full SED modeling study. \label{tab:fullCIG}}
\tablewidth{0pt}
\tablehead{
\colhead{CIGALE Module} & \colhead{Parameter} &  \colhead{Possible Values} &  \colhead{Output}   \\
}
\decimals
\startdata
{\tt\string sfhdelayed} & \mathrm{E-folding \; time \; of \; main \; stellar \; pop.} & 500, 1000, 2000, 4000 \; \mathrm{Myr} & 4000 \mathrm{\; Myr}\\
 & \textrm{Main stellar pop. age} & 5000, 7000, 9000, 11000 \mathrm{\; Myr} & 9000 \mathrm{\; Myr} \\
 & \textrm{Burst pop. mass fraction} f_{\mathrm{burst}} & 0.0, 0.3, 0.6, 0.9 & 0.0\\
\hline
{\tt\string bc03} & \mathrm{IMF} & \textrm{Salpeter, Chabrier} & \mathrm{Salpeter} \\
 & \mathrm{Metallicity} & 0.004,  0.02 & 0.02 \\
  & \mathrm{Separation \; Age} &  500, 750, 1000, 2500, 5000, 7500 & 7500 \; \mathrm{Gyr} \\
\hline
{\tt\string nebular} & \mathrm{Ionization \;} \log (U) & -2.0 & -2.0 \\
 & \mathrm{Escaped \; Lyman \; Cont. Fraction} & 0.0 & 0.0 \\
 & \mathrm{Absorbed \; Lyman \; Cont. Fraction} & 0.0 & 0.0 \\
 & \mathrm{Line\; width} & 70 \si{km/s}  &  70 \si{km/s} \\
\hline 
{\tt\string dustatt\_modified\_CF00}  & A_{\textrm{V, ISM}}& 0.1, 0.35, 0.65, 1.0 & 0.1 \\
  & A_{\textrm{V, ISM}} / \left( A_{\textrm{V, Birth Clouds}} + A_{\textrm{V, ISM}}\right) & 0.44 & 0.44 \\
  & \alpha_{\mathrm{ISM}}& -0.7 & -0.7 \\
  & \alpha_{\textrm{Birth Clouds}}& -1.3 & -1.3 \\
\hline
{\tt\string dl2014} &  \textrm{PAH Mass Frac.} & 1.12, 2.50, 3.90, 5.26, 5.95, 7.32 & 2.5 \\
 & \textrm{Min Rad. Field} \; U_{\mathrm{min}} & 0.5, 1, 2, 3, 5, 8, 10 & 5 \\
 & \textrm{Powerlaw slope } \alpha, \; (dU/dM \propto U^{\alpha}) & 1.0, 2.0, 3.0 & 3.0 \\
 & \textrm{Frac. illuminated from} \; U_{\mathrm{min}} \textrm{ to } U_{\mathrm{max}}& 0.1, 0.3, 0.5, 0.7, 0.9 & 0.7\\
 \hline 
 {\tt\string skirtor}  & \tau_{9.7 \; \si{\micro\meter}}& 9, 11 & 11 \\
 & p_{\ell}& 0.0, 0.5, 1.0, 1.5 & 0.5 \\
 & q& 0.0, 0.5, 1.0, 1.5 & 0.0 \\
 & \theta & 10, 20, 30, 40, 50, 60, 70, 80 & 70 \\
 & R_{\textrm{out}} / R_{\textrm{in}} & 10, 20, 30 & 20 \\
 & i & 0, 10, 20, 30, 40, 50, 60, 70, 80, 90 & 10 \\
 & {\tt\string fracAGN} & 0.0, 0.2, 0.4, ... 0.98 & 0.6 \\
 \hline
 {\tt\string fritz2006} & R_{\mathrm{\, ratio}} & 10, 30, 60, 100, 150 & 30 \\
 & \tau & 0.1, 0.3, 0.6, 1.0, 2.0, 3.0, 6.0, 10.0 & 10.0 \\
 & \beta & -1.00, -0.75, -0.50, -0.25, 0.00 & -0.25 \\
 & \gamma & 0.0, 2.0, 4.0, 6.0 & 6.0 \\
 & \theta & 30, 50, 70 & 70.0 \\
 & i &  0, 10, 20, 30, 40, 50, 60, 70, 80, 90 & 30   \\
 & {\tt\string fracAGN} & 0.0, 0.2, 0.4, ... 0.98 & 0.6 \\
\enddata
\end{deluxetable*}

\subsection{AGN + Nuclear Star Cluster SED Fitting}
\label{sec:totalsedmodelling}

Using the range of best-fit parameters from \S\ref{sec:galaxysed} as limits, we jointly model our full range of multiwavelength data for the combined galaxy and AGN emission. Specifically, we include the spectrophotometry extracted from the SDSS and \emph{Spitzer} spectroscopy, combined with the K-band MAGNUM measurement and the photometry extracted from the \emph{Spitzer} IRAC imaging. Since we consider the fit from the previous section to be our best guess of the true level of galaxy light, we must simultaneously fit the global photometry to the overall SED as well as the \textit{HST} photometry to the dust + stellar SEDs.

We input the best-fit galaxy parameters from above as a starting position for a global fit that solves for both an AGN and galaxy component.  We fit the galaxy parameters, the overall AGN amplitude ({\tt\string fracAGN}), and the six torus parameters; the results of this fit are shown in Figure \ref{fig:fullSED}. This is a successful fit, but does not necessarily enforce a good fit (in a $\chi^2$ sense) between the \emph{HST} data and the galaxy model. To enforce this additional constraint, we tune the galaxy amplitude manually using fracAGN, which is a measure of the fraction of light contributed by the AGN relative to the total infrared luminosity at 3--1000$\mu$m. We run X-CIGALE across fixed values of $0<{\tt\string fracAGN}<1$, fix five of the six torus parameters to their best-fit values, and leave one free for the program to fit. We repeat this process for all six torus parameters. Then, we examine the best-fit model with a {\tt\string fracAGN} that exhibits the minimum $\chi^{2}$ between the \textit{HST} photometry and the stellar + dust model SED. Four of the six torus parameters that most affect the SED are shown in Figure \ref{fig:1dchi2}.  We find that the results of this minimum $\chi^{2}$ analysis converge on the same best-fit torus parameters as the global all-free fit and to the same value ${\tt\string fracAGN} = 0.6$, exhibiting an AGN luminosity of $1.4 \times 10^{40}$~erg/s, consistent with that found previously ($\sim 1.9 \times 10^{40}$~erg/s) from considering the UV/optical SED \citep{moranetal1999}.

\begin{figure*}[t!]
    \centering
    \includegraphics[width=\textwidth]{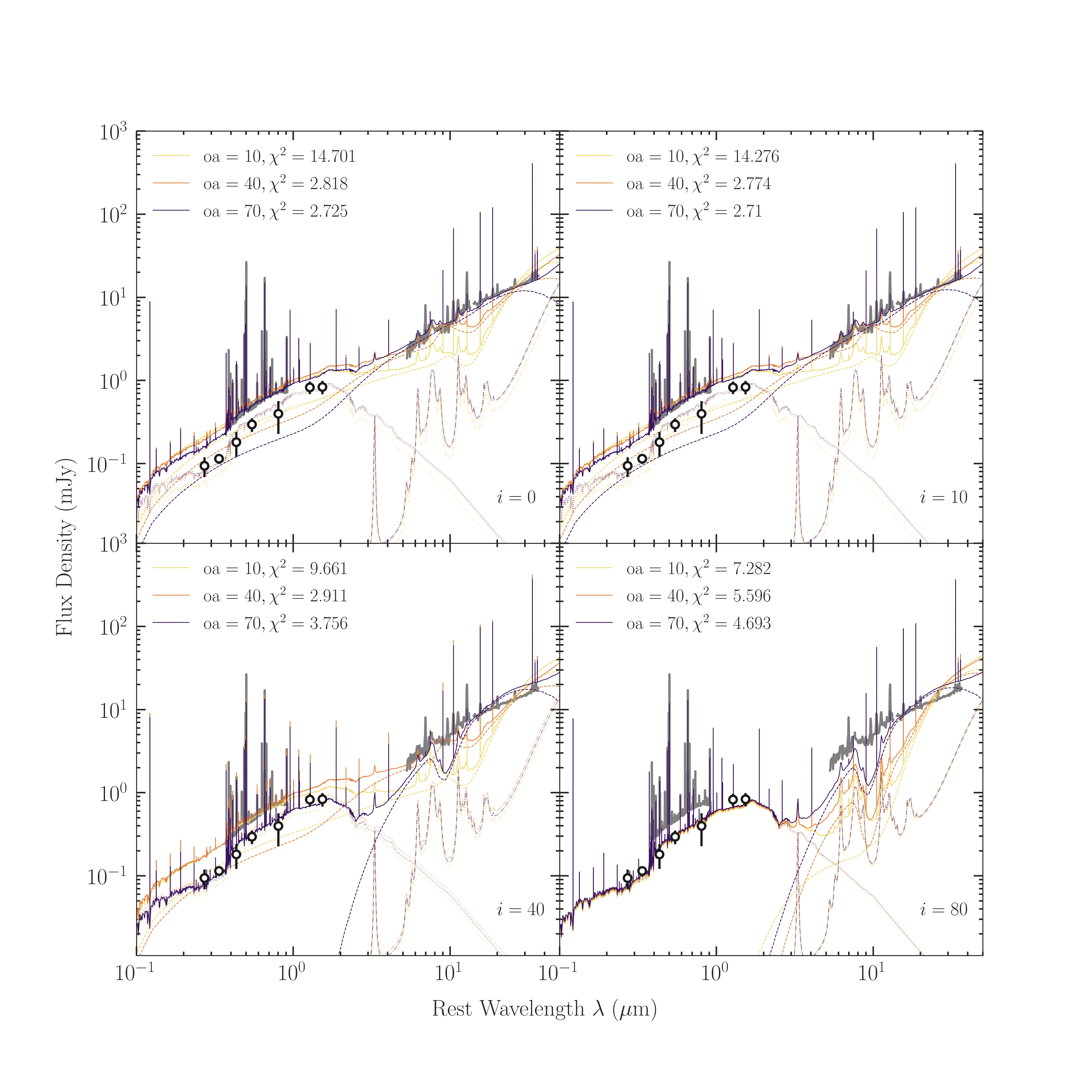}
    \caption{The different X-CIGALE model SED fits with alternating values for opening angle $\theta$ and inclination angle $i$. The SDSS and \textit{Spitzer} IRS spectrum as well as the \textit{HST/WFC3} NSC photometry have been overplotted for consistency. Note that the models demonstrate a diminishing optical component of the AGN with increasing inclination, presumably due to a line of sight passing through the obscuring torus.     }
  \label{fig:psyoa}
\end{figure*}

These additional limited free parameter fits confirm that the model with all parameters free is in satisfactory agreement with the \emph{HST} NSC data, with a reduced $\chi^{2}$ of $2.944$. The best-fit model highlights a significant contribution from starlight to the optical portion of the SED, anchored by the spectrophotometry extracted from the 3$''$ SDSS fiber. By contrast, the best-fit model suggests that there is little galaxy contribution to the MIR spectrum. We also find that the IRS spectrum is well-fit by the model; the spectrum exhibits a smooth featureless continuum, with subdominant PAH features and the absence of a silicate feature at 9.7 \si{\micro\meter}. 

To further investigate possible degeneracy between torus parameters, we then fix the best-fit model and scan over each torus parameter to examine how they affect the shape of our model SED. As silicate features in AGN SEDs are produced in the innermost BH-facing surface of the torus, the parameters most directly responsible for altering the shape of the near to mid-IR spectrum are the torus opening angle and line-of-sight inclination. We show a representative selection of the SED fits at alternating orientations and geometries in Figure \ref{fig:psyoa}. In general, we find that the photometry best favors models with wide torus opening angles and face-on inclinations. As inclination increases and the torus becomes edge-on, obscuration removes blue continuum from the AGN, pushing our fit to face-on configurations. 

\begin{figure}[t]
    \centering
    \includegraphics[width=0.49\textwidth]{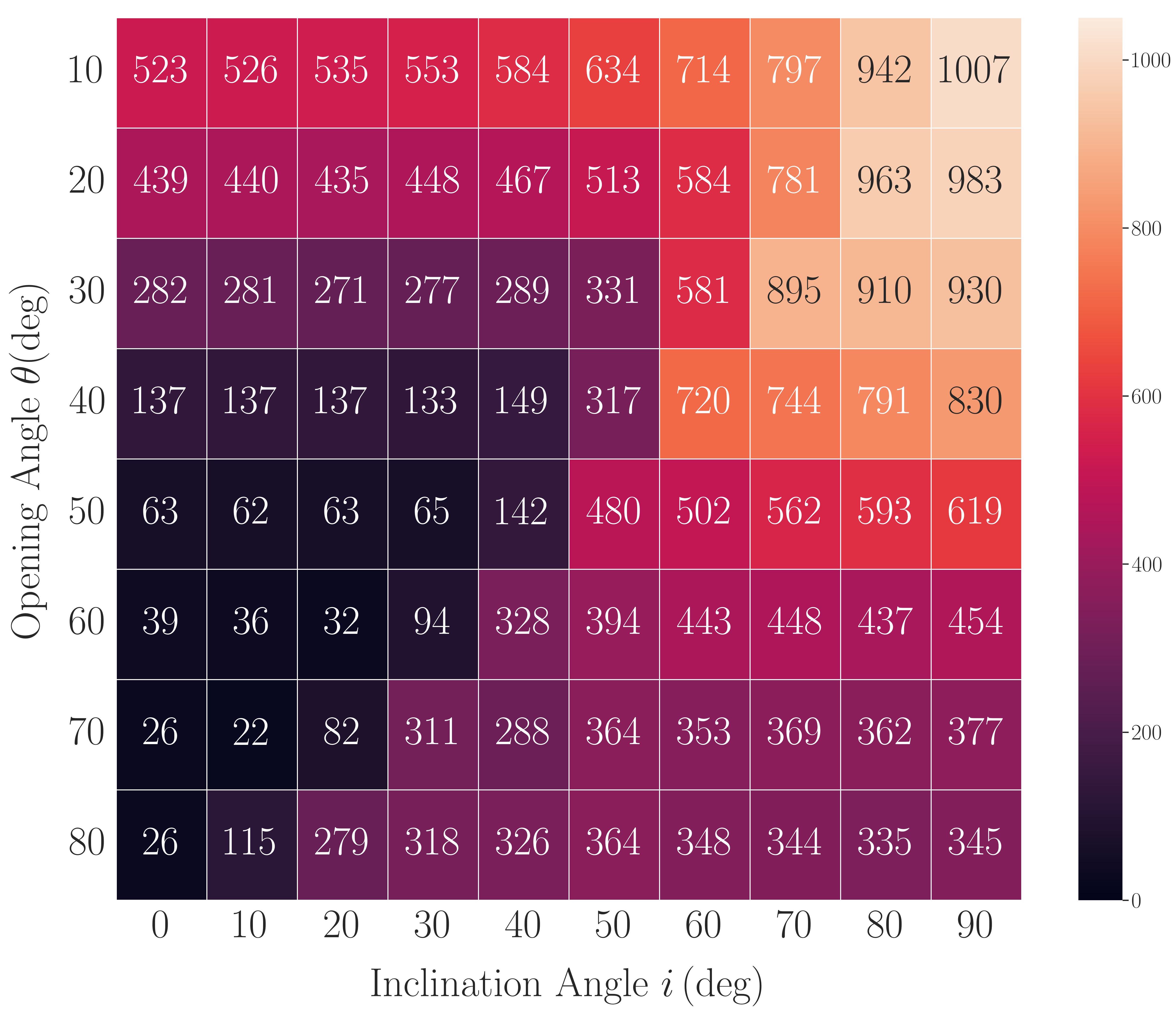}
    \vspace{-0.4cm}
    \caption{$\chi^{2}$ distribution of orientation angle ($i$) and half opening angle ($\theta$), derived from the modeled galaxy to the high spatial resolution (\hst ) described in \S\ref{sec:galaxysed}. The heatmap highlights clear preference for extremes of allowed parameters, preferring an almost perfectly face-on inclination of $10^{\circ}$ and a large half-opening angle of $70^{\circ}$. Combinations of inclination and opening angle with the line-of-sight passing through the torus are heavily disfavored.}
  \label{fig:heatmap}
\end{figure}

With this more fullsome understanding of the parameter space, we present the full range of allowed geometries and orientations in Figure \ref{fig:heatmap}, which shows the $\chi^{2}$ map of our SDSS + IRS + MAGNUM spectrophotometry relative to the overall best-fit SED model across half-opening angles $0^{\circ} < \theta < 80^{\circ}$ and inclinations $0^{\circ} < i < 90^{\circ}$. The model strongly prefers extremes of the allowed parameters, with a half opening angle of $\theta = 70 ^{\circ}$ and an inclination of $i = 10^{\circ}$ indicating a face-on configuration. Inclination and opening angle combinations looking through the torus are heavily disfavored. We conclude that the torus of NGC 4395 is very well-constrained to be (a) clumpy (b) face-on and (c) wide-angle. This model makes a strong prediction that at wavelengths beyond $\sim 20$\micron, dust emission from the galaxy will outshine dust from the torus. It would be useful to test this prediction with longer-wavelength data.

\subsection{Investigating the use of Alternative Torus Models}
\label{sec:smooth}

To this point, we have focused on the use of clumpy torus models to reproduce the observed IR emission, as these have become more common place in the recent literature, and are understood to better represent IR emission from tori, particularly when the highest spatial resolution data are available \citep[see][for reviews]{Ramos2017,honig2019}. The {\tt skirtor} model that we have employed thus far is a two-phase model using both smooth and clumpy distributions for the dust. Such two-phase models were designed in part based on the predictions of hydrodynamical simulations showing that the torus is likely a multiphase structure \citep[e.g.,][]{schartmannetal2014}. 

\begin{figure}[t]
    \centering
    \includegraphics[width=0.49\textwidth]{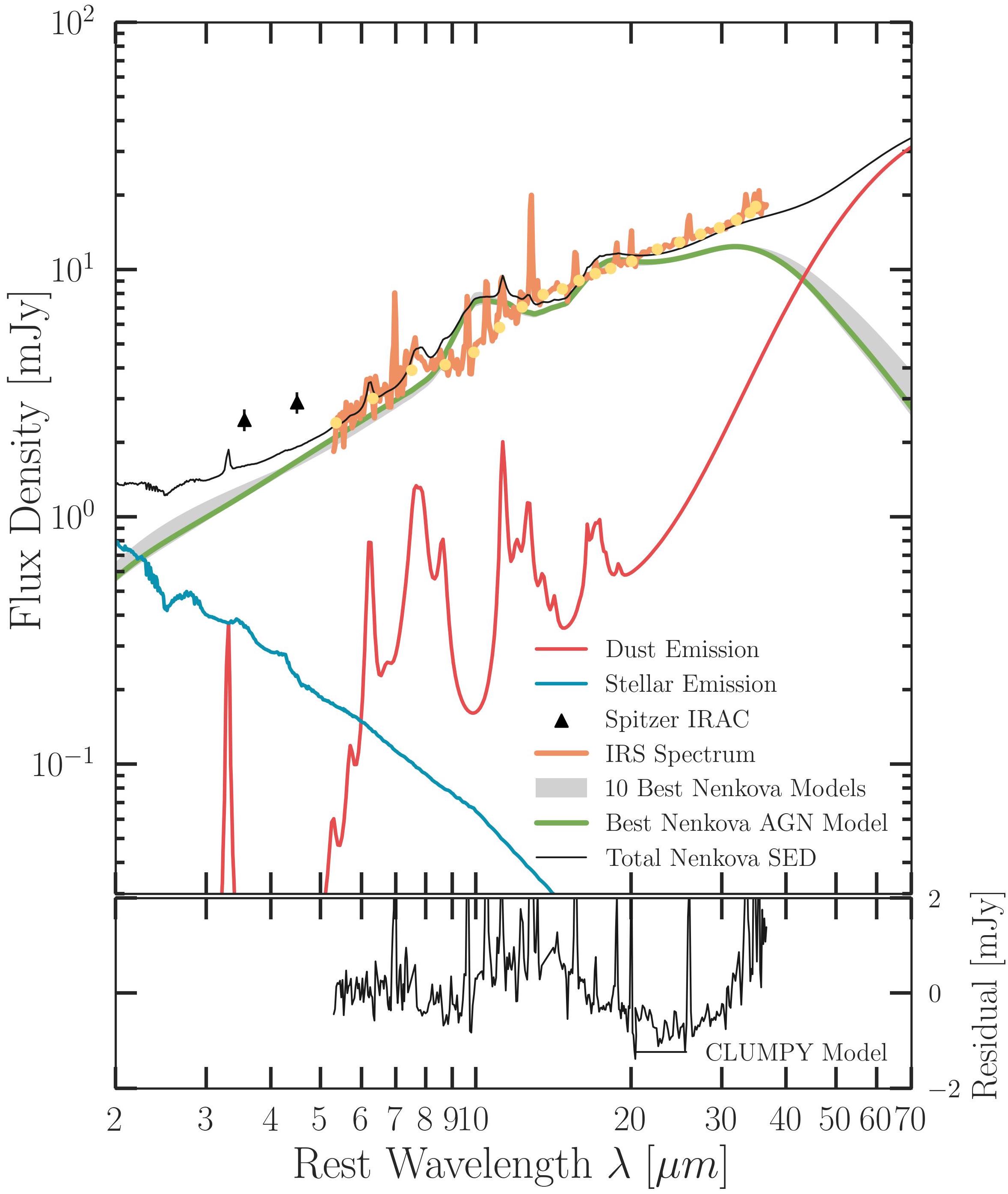}
    \vspace{-0.4cm}
    \caption{ SED comparison of the \textit{Spitzer}-IRAC photometry (black triangles) and \textit{Spitzer}-IRS spectrophotometry, against the {\tt\string CLUMPY} torus models. UV/optical stellar component and dust stellar component predicted by X-CIGALE in blue and red, respectively. In gray are the space of ten {\tt\string CLUMPY} AGN SEDs closest in fit to the IRS spectrum, with the best-fitting SED represented in green. For illustrative purposes, the \textit{Spitzer}-IRS spectrum is overlaid in orange. }
  \label{fig:nenkova}
\end{figure}
Here we replace the {\tt skirtor} model in X-CIGALE with the smooth torus models of \citet{fritzetal2006}. The \citet{fritzetal2006} model assumes a smooth flared-disk dust torus geometry, parameterized by the line-of-sight inclination (where edge-on is $\Psi = 0^o$), optical depth ($\tau$), the ratio of the maximum to minimum radii, the opening angle, and the density distribution of the dust contained within the torus. Through a radiative transfer model the output AGN SED consists of three components, primary emission from the disk, a dust scattered component, and dust re-emitted component. We perform consistent analyses to those outlined in the previous sections using this Fritz smooth torus model. The results of the fit is shown as the dashed line in Figure \ref{fig:fullSED}. Most notably, this best-fit model produces a silicate emission feature at 9.7 \si{\micro\meter}, which is in stark contrast to the observed \textit{Spitzer}-IRS spectrum, as well as the best-fit {\tt skirtor} clumpy model. Furthermore, we find increased residuals blueward of $\lambda < 20 \mu$m. We further investigate this in Figure~\ref{fig:psyoa}, where we show that for the smooth torus model fits, only face-on inclinations provide the necessary blue UV/optical continua for the AGN, but in turn produce strong Si-emission features that are not observed in the IR. By contrast, even moderately inclined tori, that reduce the Si-feature produce little to no UV continua. Thus, we conclude that a smooth torus alone cannot simultaneously provide the needed blue/UV light from a face-on torus and accommodate the lack of Si emission in NGC~4395. Only clumpy torii with high covering fractions can simultaneously yield negligible Silicate emission or absorption and the blue continuum of an unobscured AGN.

We also conduct a complementary investigation of strictly clumpy models from the {\tt\string CLUMPY} family of torus SEDs \citep{nenkova1, nenkova2}. The {\tt\string CLUMPY} model assumes a heterogenous distribution of dusty clouds quantified similarly as the {\tt skirtor} models. The added relevant parameters control the average number of clouds along radial equatorial rays $N_0$, and the torus thickness parameter $\sigma$, the latter of which is the most analogous to the half-opening angle $\theta$. Through a radiative transfer model based on the {\tt\string DUSTY} code \citep{dusty5, dusty4, dusty3, dusty2, dusty1}, the output AGN SED consists of a torus component and an input AGN spectrum. We fit the \textit{Spitzer}-IRS spectrophotometry against the {\tt\string CLUMPY} SED models while fixing the stellar and dust components to the SED to those suggested from our forward-modeling process in Sections \ref{sec:galaxysed} and \ref{sec:totalsedmodelling}. 

We present the region occupied by the ten closest-fitting models in shaded gray, alongside the best-fitting model in green in Figure \ref{fig:nenkova}. The best-fit {\tt\string CLUMPY} model represents a torus of inclination $i = 50^o$, angular torus thickness $\sigma = 25^o$, and radial extend $Y \equiv R_\mathrm{out}/R_\mathrm{in} = 100$. Similar to the {\tt\string fritz} models, the closest {\tt\string CLUMPY} models all insist on a silicate emission feature at 9.7 \si{\micro\meter}, in disagreement with the observed \textit{Spitzer}-IRS spectrum as well as the best-fit {\tt skirtor} clumpy model. From analyses of intermediate-type AGN with {\tt\string CLUMPY} torus models in the literature \citep{garciabernete19}, NGC 4395 shows similar inclination angles to those of more luminous Type 1.5 Seyferts. However, these external {\tt\string CLUMPY} fits show incomparable angular and radial extents and disagreement on the presence of silicate features.  Overall, analyses invoking {\tt\string CLUMPY} torus models produce a wide variety of potential radial extents and predict low inclination angles, while simultaneously predicting narrow opening angles \citep{ramosalmeida09, alonsoherrero11, audibert17}. The latter is in contrast to the predictions using the {\tt\string skirtor} model. From the overall geometry and lack of prominent silicate features predicted by the {\tt\string skirtor} fits, we find that their two-phase torus models are the most proximate in describing the SED of the AGN in NGC 4395.

\subsection{The Incompleteness of the Clumpy Torus Model}

While the two-phase clumpy torus model does a better job of reproducing the lack of Si emission or absorption with the steep mid-infrared slope compared with a simpler smooth torus, the fit is not particularly good in the 3-6 \micron\ region of the spectrum. This is a known deficiency of clumpy torus models \citep[e.g.,][]{honigetal2017,garciagonzalezetal2017,gonzalezmartinetal2019}. In particular, \citet{honigetal2017} argue that to account for the additional near-IR emission, a second torus component is required. This second component is seen in interferometric observations of some nearby AGN \citep[e.g.,][]{burtscheretal2013,lopezgonzagoetal2016}. \citet{honigetal2017} present torus models that explicitly include a poloidal wind component, and argue that the wind emission dominates the mid-infrared emission, while a blue near-IR spectral slope (the so-called blue bump) is dominated by the outer accretion disk. An example of such a ``wind'' model is implemented in the {\tt CAT3D-WIND} SED library. 

We follow a similar methodology to that implemented previously \citep[e.g.,][]{hernancaballeroetal2015,garciagonzalezetal2017,honigetal2017} and perform a non-parametric test to elucidate if a wind model can reproduce the observed mid-IR parameters in NGC~4395. For consistency with these previous studies, we use the {\tt DeblendIRS} package \citep{hernancaballeroetal2015} to empirically decompose the {\it Spitzer}-IRS spectra into its `pure-AGN', `pure-stellar', and `pure-interstellar' sub-components. From the AGN component, we then measure the near-IR and mid-IR spectral slopes ($\alpha_{\rm NIR}$ and $\alpha_{\rm MIR}$ respectively) and the equivalent width of the Silicate feature.

    \begin{figure*}[t]
    \centering
        \includegraphics[width=0.85\textwidth]{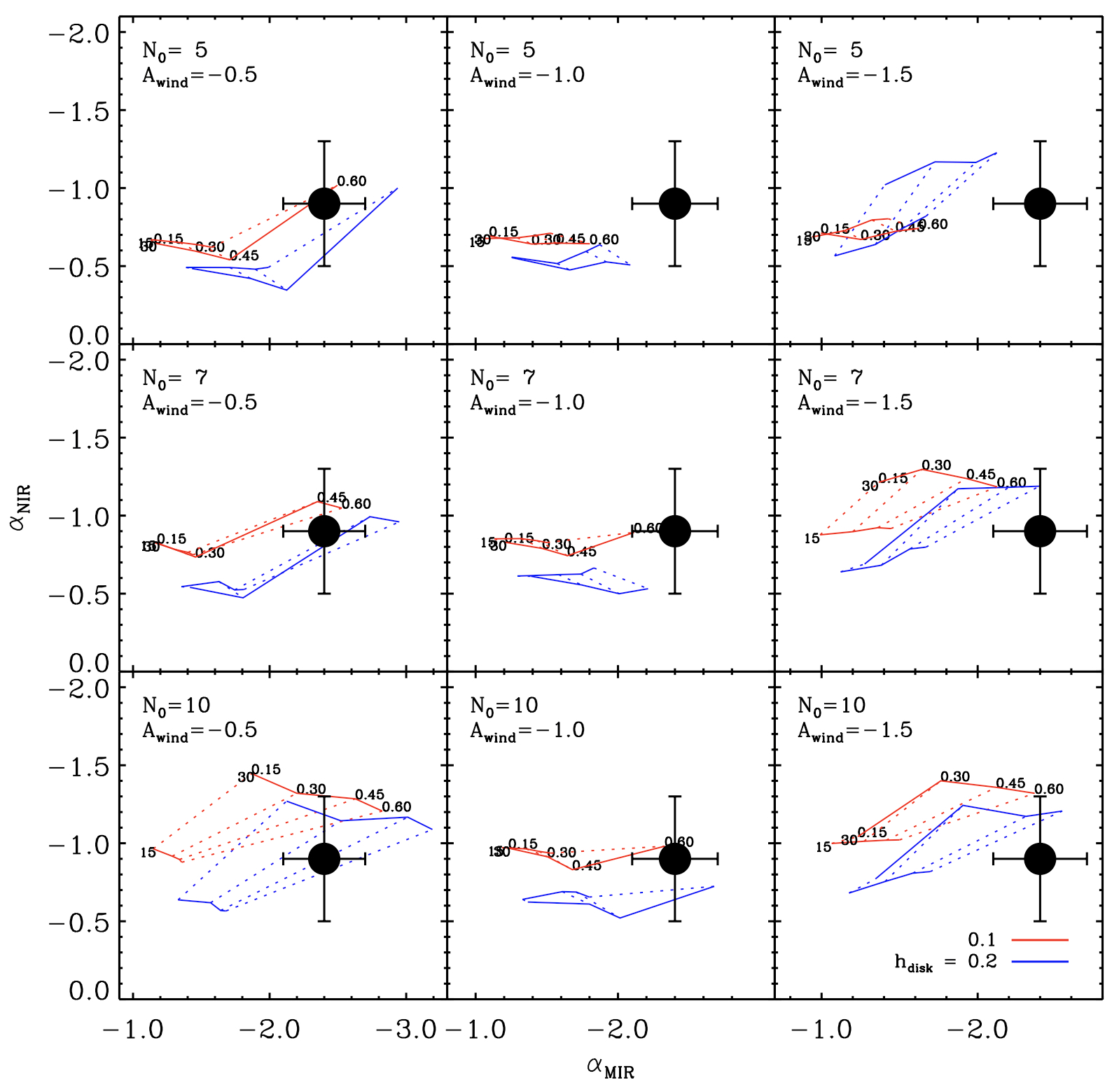}
    \caption{Non-parametric plot of mid-IR ($\alpha_{\rm MIR}$) and near-IR ($\alpha_{\rm NIR}$) spectral slopes of the AGN emission in NGC~4395 (filled circle). For comparison, {\tt CAT3D-WIND} models are shown for 15$^o$ and 30$^o$ inclination angles (solid lines), varying fractions of the polar wind contribution to the mid-IR emission ($f_{\rm wind} = \{0.15, 0.30, 0.45, 0.60\}$; dashed lines), and disk scale heights of 0.1 and 0.2 (red, blue lines, respectively). Panels provide number of dust clumps along the line of sight ($N_0 = \{5, 7, 10\}$) and the radial distribution of dust clouds in the wind ($A_{\rm wind} = \{-0.5, -1.0, -1.5\}$). All models shown assume a radial power law slope of $a = -2.5$, and a half-opening angle and angular-width of the wind $\theta_W = 45^o$ and $\sigma_{\theta} = 15^o$, respectively, see \citet{honigetal2017} for details of the {\tt CAT3D-WIND} models.}
  \label{fig:CAT3DGrids}
\end{figure*}

In Figure \ref{fig:CAT3DGrids}, we show NGC 4395 in $\alpha_{\rm NIR}$ and $\alpha_{\rm MIR}$ space. We compare with a range of models produced in {\tt CAT3D-WIND}, with relatively thin (scale heights of 0.1 and 0.2) face-on disks with inclination angles of 15$^o$ and 30$^o$ (similar to those found in our earlier analyses), and varying fractions of the polar wind contribution to the mid-IR emission ($f_{\rm wind} = \{0.15, 0.30, 0.45, 0.60\}$). We include models with a varying number of dust clumps along the line of sight ($N_0 = \{5, 7, 10\}$) and changing radial distribution of dust clouds in the wind ($A_{\rm wind} = \{-0.5, -1.0, -1.5\}$). For all models, we assume a radial power law slope of $a = -2.5$, and a half-opening angle and angular-width of the wind $\theta_W = 45^o$ and $\sigma_{\theta} = 15^o$, as these produced the most meaningful results based on the NGC~4395 measurements.

We find that it is possible to explain the location of NGC~4395 in $\alpha_{\rm NIR}$--$\alpha_{\rm MIR}$ space with reasonable {\tt CAT3D-WIND} models, particularly those invoking high $f_{\rm wind}$ fractions. Interestingly, the NGC~4395 measurements are consistent with the region in the {\tt CAT3D-WIND} models where the Si absorption transitions to emission, which is also consistent with our observations. The level of Si absorption ($-S_{\rm Si} \sim 0.1$--0.2) observed in NGC~4395 is also similar to the models and data presented in \citet{garciagonzalezetal2017} for other Type 1.8/1.9 AGN. We further note that the $\alpha_{\rm MIR}$ slope is significantly steeper than $\alpha_{\rm NIR}$ owing to the existence of the NIR blue blump; this feature cannot be reproduced by the no-wind CAT-3D models. \citet{honigetal2017} use the existence of AGN in the space where $\alpha_{\rm NIR}$ is bluer than $\alpha_{\rm MIR}$ to argue that the wind model is needed to describe real AGN. This need for a wind in some sources is also in accord with interferometric observations of a small handful of nearby Seyfert galaxies, where a polar component is imaged in the torus-emitting region \citep[e.g.,][]{Raban:2009,Honig:2012,Tristram:2014,Leftley:2018}.

Our measurements of NGC~4395 throughout this study are in broad agreement with those most recently found for the {\tt CAT3D-WIND} and smooth torus models of this galaxy by \citet{GarciaBernete:2022}, albeit that \citet{GarciaBernete:2022} utilize only mid-IR data from 8-10m-class ground-based telescopes for their study. By contrast, the best-fit two-phase {\tt skirtor} model in \citet{GarciaBernete:2022} is that of a completely edge-on torus, which is inconsistent with our findings here. We do not have a full explanation for this difference, as the SED shapes reported by them are similar to those considered here. At the same time, it is important to note that \citet{GarciaBernete:2022} do not use the results for NGC~4395 in their wider study of hard X-ray detected AGN, presumably due to the inconsistency they find between the different model parameters and overall poor fits for NGC~4395. Consistent with our findings here, \citet{GarciaBernete:2022} also show that the {\tt CAT3D-WIND} model produces the best overall fit to NGC~4395 when the near-IR data are also considered in their fitting.

\section{Discussion \& Conclusion} \label{sec:discussion}

In the previous sections, we determined that the preferred geometry for the obscuring material surrounding the central BH in NGC 4395 involves a semi-coherent structure of many individual optically thick clouds \citep{nenkovaetal2002,nenkovaetal2008}, a so-called clumpy torus model that we parameterized using the {\tt skirtor} library in X-CIGALE. Due mainly to the differing line-of-sight heating and radiative transfer effects of the clouds, these models tend to produce less pronounced Silicate-absorption features for accretion-disks that are viewed edge-on, and a wider range of mid-IR spectral slopes. 

Over the past decade, the idea of a polar-wind component to the torus has also gained support, both due to spatially resolved observations of a poloidal component to the torus \citep[e.g.,][]{Tristram:2014} and due to the success of dusty wind models in reproducing the NIR+MIR SEDs of AGN \citep[e.g.,][]{garciagonzalezetal2017,GarciaBernete:2022}. Theoretically, we do expect that radiation pressure can drive winds in dusty environments \citep[e.g.,][]{Thompson2015}. Recently, \citet{Venanzi:2020} presented simulations of radiation-pressure driving from dusty disks in the torus region of AGN. They show that the potential for driving a wind will depend on $N_H$, which scales with the amount of material available to be driven by the wind, and the Eddington ratio, which encodes how much radiation pressure is available to act against the gravity of the AGN. There is a sweet spot where $N_H$ is moderate (log $N_{\rm H} \approx 22-23$~cm$^{-2}$, and the Eddington ratio is relatively high $\sim 10\%$, where they expect the radiation pressure is strong enough to drive a wind.

Interestingly, \citet{GarciaBernete:2022} find some support for this picture by fitting the SEDs of tens of local AGN with NIR+MIR spectroscopy. They find the dusty wind model to be a better fit to moderately obscured AGN (e.g., Seyfert Type 1.5-1.8) and usually not required in Type 2 (more heavily obscured) systems. NGC~4395, which is classified as a Seyfert Type 1.8, and through careful modeling of its X-ray spectral shape, has been determined to show evidence for partial covering by variable cold absorbers producing a log$N_{\rm H} \sim 22-23$~cm$^{-2}$ \citep{nardinirisaliti2011,kammounetal2019}, and hence, seems to fit into this picture rather nicely. 

Our current study is still limited by two main issues: (1) the lack of simultaneity across all of the wavebands and (2) the spatial/spectral resolution of the mid-IR data out beyond 20$\mu$m. However, with the advent of the \emph{James Webb Space Telescope} (\emph{JWST}), we are now in a position to obtain a much higher resolution view of the torus region in NGC 4395, which hopefully can serve as a template for future searches.
With its sub-arcsecond resolution, JWST will be a powerful tool in accurately decomposing IR spectra where AGN light dominates into actual AGN signal and its stellar and galactic contaminants. High-resolution studies of the NIR may provide insight into possible winds, silicates, and graphites in the torus, whereas observations at longer MIR wavelengths will determine whether there is a luminous star forming feature as predicted by CIGALE. The Mid-InfraRed Instrument (MIRI) should be able to measure the hottest bit of this dusty continuum, place limits on possible PAH features, and constrain silicate features apart from any spectral contaminants. These studies of low-luminosity AGN spectral features will facilitate future surveys searching for IMBHs, shed light on the processes that allow MBHs to form, and in turn gain insight into the complexities of galaxy evolution.

\facilities{SDSS, HST, IRSA, Spitzer (IRAC, IRS)}
\software{GALFIT \citep{pengetal2002, pengetal2010}, astropy v3.1.1 \citep{astropy:2013, astropy:2018, astropy:2022}, Matplotlib \citep{hunter07}, Numpy v1.25 \citep{harris2020array}, Photutils v0.6; \citep{larry_bradley_2022_6825092}, DeblendIRS \citep{hernancaballeroetal2015}, CAT3D-WIND \citep{honigetal2017}, X-CIGALE \citep{yangetal2020, nolletal2009, serraetal2011}}

\acknowledgments
The authors thank Sebastian Honig, Anil Seth, and Luis Ho for useful discussions. This research has made use of the NASA/IPAC Infrared Science Archive, which is funded by the National Aeronautics and Space Administration and operated by the California Institute of Technology. This work also makes use of data from SDSS-II. Funding for the SDSS and SDSS-II has been provided by the Alfred P. Sloan Foundation, the Participating Institutions, the National Science Foundation, the U.S. Department of Energy, the National Aeronautics and Space Administration, the Japanese Monbukagakusho, the Max Planck Society, and the Higher Education Funding Council for England . The SDSS is managed by the Astrophysical Research Consortium for the Participating Institutions. The Participating Institutions are the American Museum of Natural History, Astrophysical Institute Potsdam, University of Basel, University of Cambridge, Case Western Reserve University, University of Chicago, Drexel University, Fermilab, the Institute for Advanced Study, the Japan Participation Group, Johns Hopkins University, the Joint Institute for Nuclear Astrophysics, the Kavli Institute for Particle Astrophysics and Cosmology, the Korean Scientist Group, the Chinese Academy of Sciences (LAMOST), Los Alamos National Laboratory, the Max-Planck-Institute for Astronomy (MPIA), the Max-Planck-Institute for Astrophysics (MPA), New Mexico State University, Ohio State University, University of Pittsburgh, University of Portsmouth, Princeton University, the United States Naval Observatory, and the University of Washington.

Some/all of the data presented in this paper were obtained from the Mikulski Archive for Space Telescopes (MAST) at the Space Telescope Science Institute. The specific observations analyzed can be accessed via \dataset[https://doi.org/10.17909/pw8k-wb82]{https://doi.org/10.17909/pw8k-wb82}. STScI is operated by the Association of Universities for Research in Astronomy, Inc., under NASA contract NAS5–26555. Support to MAST for these data is provided by the NASA Office of Space Science via grant NAG5–7584 and by other grants and contracts.

\newpage

\bibliography{main.bib}

\end{document}